\documentclass[12pt]{iopart}
\usepackage{graphicx,color} % for pdf, bitmapped graphics files
\usepackage[latin1]{inputenc}
\usepackage[OT1]{fontenc}
\usepackage[english]{babel} % for proper hyphenation patterns
\usepackage{booktabs,multirow,times}
\usepackage{float} % para usar [H] 
\usepackage[right]{rotating}
\usepackage[abbr]{harvard}
\begin{document}

\title[The SWT as an Efficient Reductor of PLI in Cardiac Electrophysiology]{\Large The Stationary Wavelet Transform as an Efficient Reductor of Powerline Interference for Atrial Bipolar Electrograms in Cardiac Electrophysiology}

\author{ Miguel Mart\'inez-Iniesta$^{1}$, Juan R\'odenas$^1$, Jos\'e J. Rieta$^2$ and Ra\'ul Alcaraz$^1$}
\address{$^1$ Research Group in Electronic, Biomedical and Telecommunication Engineering,
University of Castilla-La Mancha, Spain.}
\address{$^2$ BioMIT.org, Electronic Engineering Department, Universitat Politecnica de Valencia, Spain.}
\ead{\mailto{miguel.martinez@uclm.es}}%, \mailto{juan.rodenas@uclm.es}, \mailto{jjrieta@upv.es}  and \mailto{raul.alcaraz@uclm.es}}

%%%%\date{}
%%%
%%%%\linenumbers
%%%%\maketitle

\begin{abstract}
\emph{Objective:} The most relevant source of signal contamination in the cardiac electrophysiology (EP) laboratory is the ubiquitous powerline interference (PLI). To reduce this perturbation, algorithms including common fixed-bandwidth and adaptive-notch filters have been proposed. Although such methods have proven to add artificial fractionation to intra-atrial electrograms (EGMs), they are still frequently used. However, such morphological alteration can conceal the accurate interpretation of EGMs, specially to evaluate the mechanisms supporting atrial fibrillation (AF), which is the most common cardiac arrhythmia. Given the clinical relevance of AF, a novel algorithm aimed at reducing PLI on highly contaminated bipolar EGMs and, simultaneously, preserving their morphology is proposed. \emph{Approach:} The method is based on the wavelet shrinkage and has been validated through customized indices on a set of synthesized EGMs to accurately quantify the achieved level of PLI reduction and signal morphology alteration. Visual validation of the algorithm's performance has also been included for some real EGM excerpts. \emph{Main results:} The method has outperformed common filtering-based and wavelet-based strategies in the analyzed scenario. Moreover, it possesses advantages such as insensitivity to amplitude and frequency variations in the PLI, and the capability of joint removal of several interferences. \emph{Significance:} The use of this algorithm in routine cardiac EP studies may enable improved and truthful evaluation of AF mechanisms.
\end{abstract}
 
\submitto{\PM}
\maketitle

\section{Introduction} \label{sec:intro}
The acquisition of biomedical recordings is usually corrupted by the presence of noise and nuisance interferences~\cite{Schanze2017}. In addition to the internal noise introduced by the recording system as well as common baseline wandering from patient's respiration, other disturbances from muscular activity of the patient and the powerline interference (PLI) often reduce significantly the quality of these signals~\cite{Metting1990,Venkatachalam2011b}. This last perturbation mainly consists of a sinusoidal signal centered on 50/60~Hz, which can present amplitude and frequency variations, as well as a moderate content of harmonic distortion~\cite{Warmerdam2017}. For its reduction, current recording systems include shielded wires as well as amplifiers with high common mode rejection ratios. However, these protective measures are insufficient and the PLI still disturbs a variety of both surface and invasive recordings, including electrocardiograms (ECGs), electroencephalograms (EEGs), intra-cardiac electrograms (EGMs) and intra-cranial electrocorticograms (EoCG)~\cite{Warmerdam2017}. 
\par
As a consequence of such recording situation, the accurate characterization of the physiological information captured in biomedical signals requires the development of effective preprocessing algorithms for noise removal. However, this is a major challenge, since the PLI falls within the bandwidth of interest for most physiological recordings, i.e., within the frequency range 1--150~Hz for the ECG, 0.1--120~Hz for the EEG, 1--300~Hz for the EGM and 0.1--200~Hz for the EoCG~\cite{Webster2014,Venkatachalam2011a}. Hence, PLI removal demands a strong trade-off between rejection of the unwanted noise and preservation of the original signal morphology~\cite{Allen2009}. In fact, an extensive presence of PLI as well as large alteration of the original waveform have been identified as potential sources of misdiagnosis and misapplication of clinical treatments under different scenarios of ECG recording~\cite{Bakker2010,Zivanovic2013,Bcharri2017}. Moreover, both situations are especially detrimental in the analysis of intra-atrial EGM recordings, which are the basis to understand atrial fibrillation (AF), because its mechanisms are still under intensive research and are not completely known~\cite{Schotten2016,Lau2016}.
\par
More precisely, bipolar EGMs capture local electrical activity directly on the cardiac tissue, thus providing precise information about time, direction and complexity of local atrial activations (LAAs)~\cite{Stevenson2005}. These signals are therefore the best source of information to improve current knowledge about the particular mechanisms triggering and sustaining AF in every patient. Nonentheless, although EGMs have been widely characterized in terms of rate, organization, and morphology~~\cite{Ravelli2014,Heijman2016,Baumert2016}, a meaningful progress in their interpretation has been recently highlighted as a priority, because AF is one of the major cardiovascular problems in developed countries~\cite{VanWagoner2015}. Indeed, AF is the most common cardiac arrhythmia in clinical practice, affecting approximately  1.5--2\% of the general population~\cite{ZoniBerisso2014}. Furthermore, since its prevalence is closely related to age and elderly population is  growing very quickly, it is expected that the number of patients suffering from AF vastly increases in coming years. Thus, some recent epidemiological studies have suggested that this arrhythmia will reach epidemic proportions by the middle of this century~\cite{Krijthe2013,Colilla2013}.  
\par
Moreover, it should also be remarked that the current first-line therapy to treat AF patients, i.e. catheter ablation, also relays heavily on intensive analysis of EGM recordings~\cite{Goldberger2017}. To this respect, beyond the successful isolation of the pulmonary veins for patients in an initial stage of AF, extensive ablation has been widely proposed for patients suffering from a persistent arrhythmia~\cite{Romero2017}. However, in this case the optimal set of atrial lesions is still to be determined, because conflicting results have been reported for most alternatives proposed to date~\cite{Quintanilla2016}. Anyway, these ablation protocols are based on targeting atrial regions which supposedly are AF sustainers and these targets can be exclusively identified by applying intensive signal processing in frequency and morphology domains to intra-atrial EGM recordings. Hence, successful and efficient PLI removal plays a key role, because large residual or alteration of the EGM morphology may lead to targeting wrong atrial zones, thus limiting the success rate of those procedures~\cite{Bakker2010}.  
\par
Despite this context, PLI suppression from EGM recordings has not received enough attention in the last years. Unlike the ECG, in which a broad variety of PLI denoising algorithms have been proposed~\cite{Sharma2018,Warmerdam2017,Liu2018}, this interference is mainly reduced in the EGM by using fixed-bandwidth or adaptive notch filters. These approaches have proven to add artificial fractionation to the EGM, thus concealing high-frequency activity generated from pulmonary vein deflections and low-amplitude near-field activity from other atrial structures~\cite{Jadidi2013,Venkatachalam2011a}. Indeed, many commercial recording systems only include a simple notch filter for PLI reduction~\cite{Bakker2010} and no other algorithms can be found in the literature for that purpose. Hence, to the best of our knowledge, this work introduces for the first time a novel methodology capable of reducing high levels of PLI from bipolar EGM recordings and, simultaneously, preserving their original morphology. The method is based on the well-known wavelet shrinkage approach, which has reported significantly advantageous features compared with common notch filters~\cite{Poornachandra2008}.

\section{Methods}\label{sec:methods}
\subsection{Generation of clean EGM recordings}\label{sec:gen_EGM}
\begin{figure*}[tb!]
\centering
\includegraphics[width=\textwidth,keepaspectratio]{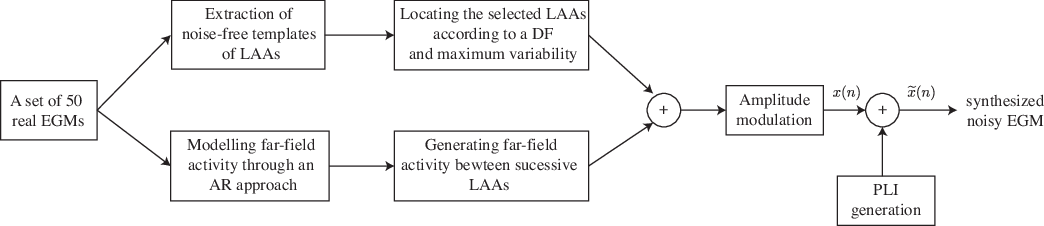}
\caption{Block diagram describing how clean EGM recordings are generated. In short, noise-free templates of LAAs are first extracted from real EGM signals. On the other hand, background activity between LAAs in real EGMs is modeled as an autoregressive (AR) approach. Then, the EGM is synthesized by locating LAAs according to a specific DF and adding a simulated background activity. Finally, the EGM amplitude is modulated according to a real signal and PLI is added with a specific SIR value.}\label{fig:gen_EGM}
\end{figure*}

To accurately quantify the achieved level of PLI reduction as well as signal morphology alteration, the proposed algorithm was validated on a set of synthesized bipolar EGM recordings. These signals were generated making use of a model previously published~\cite{Oesterlein2015,Martinez2017}. Briefly, as Figure~\ref{fig:gen_EGM} summarizes, each clean EGM was synthesized as the sum of two components, i.e., LAAs along with the far-field effect. Regarding the first item, LAAs were obtained from 50 real EGMs, which were acquired under adenosine administration to avoid far-field ventricular contamination~\cite{Atienza2006}. These excerpts were carefully selected from clean EGM intervals and they were also typically preprocessed making use of a band-pass filtering between 0.5 and 500~Hz~\cite{Venkatachalam2011a}. Two experts annotated all LAAs, so that each one was delineated by a window of 90~ms centered on its maximum slope point and normalized in amplitude. Ensembles of at least 5 activations, featured by a correlation greater than 95\%, were constructed and averaged, thus obtaining 65 noise-free templates. These waveforms were then used to mimic the atrial activity component of one hundred 10 second-length bipolar EGM recordings with a sampling rate of 1000~Hz. Precisely, for each signal a template was randomly selected as a dominant pattern and the remaining ones with higher correlation were more probable to be also chosen, according to a gaussian probability distribution. The picked waveforms were finally distributed in time for reaching a randomly determined dominant frequency (DF) in the range of 3 to 12~Hz, with a maximum variation between successive LAAs of $\pm$~25~ms.  

%%%%%%%%%%%%%%%%%%%
%%%%%%%%%%%%%%%%%%%

On the other hand, the far-field activity was synthesized by modeling through an autoregressive (AR) Yule-Walker approach the intervals between successive LAAs from the aforementioned real EGM recordings~\cite{Corino2013}. The optimal AR model order for each interval was obtained from the range 1 to 40 making use of the Akaike's information criterion~\cite{Corino2013}. The values obtained in this way for all the segments of a signal were then averaged and the resulting order was used to re-compute AR fitting for all the intervals. The mean AR coefficients were finally used to generate a background signal, which was combined with the LAAs to synthesize a clean pseudo-real EGM signal, referred to as $x(n)$. As a last step, the instantaneous amplitude of the signal was modulated according to upper and lower envelopes estimated by cubic spline interpolation from real EGM recordings. 

\subsection{Synthesis of the PLI and noisy EGM recordings}
The synthesized EGM signals were next corrupted by a simulated PLI to obtain noisy recordings, referred to as $\widetilde{x}(n)$ (see Figure~\ref{fig:gen_EGM}). In Europe the standard EN-50160 defines the main voltage characteristics for public distribution systems~\cite{Cenelec:1999tn}. Thus, power supply voltage is primarily established at 230~V and 50~Hz with maximum fluctuations about $\pm$~10\% and $\pm$~1\%, respectively. This main component can also present harmonics and inter-harmonics, but their relative power content is limited. For instance, the standard confines the relative power of the first four exact components of 50~Hz to 2, 5, 1 and 6\%, respectively, and to 0.2\% for the remaining inter-harmonics~\cite{Cenelec:1999tn}. All these aspects were considered to mimic a PLI as realistic as possible. Thus, random variations within the mentioned intervals were considered for amplitude and frequency of the main component of 50~Hz, as well as for its first four harmonics. Note that the five components were also modulated in frequency with a maximum deviation of 0.5~Hz to introduce some low-amplitude inter-harmonics. The resulting signal was named common PLI and was used to obtain noisy EGM recordings with signal-to-interference ratios (SIR) of 25, 20, 15, 10 and 5~dB.     
\par
Nonetheless, the PLI observed in many physiological recordings, including bipolar EGMs, often presents amplitude and frequency variations, as well as an harmonic content, beyond the limits established by the standard EN-50160~\cite{Zivanovic2013,Levkov2005,Bakker2010}. An unstable and nonlinear behavior of some electronic equipments, such as transformers, lamps, etc., could explain this aberrant interference~\cite{Costa2009}. Thus, the proposed algorithm was also validated on some extended conditions from the common PLI. More precisely, a sudden change in the PLI amplitude was simulated by reducing abruptly SIR from infinite to a specific value in the third second of the signal~\cite{Warmerdam2017}. Moreover, a sinusoidal variation, with a frequency between 0.5 and 2~Hz, was also considered in the PLI amplitude for the remaining 7 seconds. On the other hand, the presence of harmonics stronger than those described by the standard EN-50160 was also analyzed by generating a PLI with harmonic distortion between 5 and 10 times the common one. As an additional experiment, high frequency deviations of $\pm$~3~Hz were introduced in the main component of 50~Hz and its harmonics~\cite{Zivanovic2013}. In the last experiment, a real PLI acquired during a catheter ablation procedure was also used to corrupt clean, synthesized EGM recordings. As before, SIR values of 25, 20, 15, 10 and 5~dB were considered for all these tests. 
\par
Finally, some excerpts from real EGM recordings were also selected for their analysis. More precisely, they were chosen from signals collected by the \emph{Intracardiac Atrial Fibrillation Database} (IAFDB), which is freely available at PhysioNet~\cite{Goldberger2000}. Given that these recordings are often disturbed by different kinds of interferences, excerpts without baseline wandering and high-frequency noise were extracted. Nonetheless, all of them presented a notable PLI, as well as different degrees of fractionation.

\subsection{Stationary wavelet shrinkage for PLI reduction}\label{sec:shrinking}
\begin{figure*}[tb!]
\centering
\includegraphics[width=\textwidth,keepaspectratio]{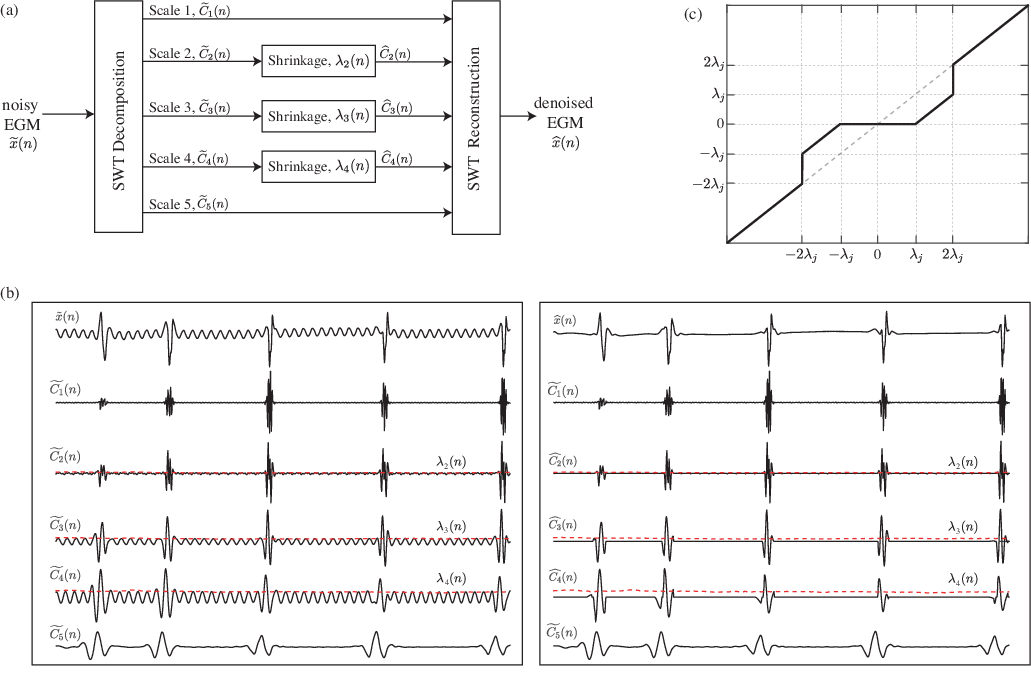}
\caption{(a) Block diagram displaying the main steps for the proposed SWT-based algorithm. (b) Wavelet coefficients for a five-level decomposition of a noisy EGM recording (left panel) and their corresponding denoised version (right panel). Thresholds $\lambda_j(n)$ are indicated as a red dashed line. (c) Transfer function for the proposed shrinkage approach based both on soft and hard thresholding functions}\label{fig:methods}
\end{figure*}

The Wavelet transform (WT) produces a time-frequency decomposition of the signal under analysis, providing a sparse representation of its information that can be splitted into a useful part and the nuisance interferences~\cite{Addison2017book}. This feature makes the WT particularly suitable for denoising of many signals, including physiological recordings~\cite{Bcharri2017,Mamun2013,Naseri2012}. In fact, if a proper orthogonal function is chosen as mother wavelet, sharp transition details of the input signal are represented as large wavelet coefficients in the vicinity of transitions and, contrarily, coefficients associated to nuisance perturbations are spread throughout the scales~\cite{Xu2004}. Then, by zeroing low-amplitude coefficients in the proper wavelet scales, noise can be mostly removed without modifying substantially the original morphology of the signal~\cite{Xu2004}. This approach, called wavelet shrinkage, has been used in the present work to reduce PLI in bipolar EGM recordings, such as Figure~\ref{fig:methods}(a) displays.  
 \par

%%%%%%%%%%%%%%%%%%%
%%%%%%%%%%%%%%%%%%%

More precisely, the noisy EGM was first decomposed into five wavelet levels by using a stationary WT (SWT). In contrast to the common discrete WT (DWT), this kind of transform is able to maintain the original resolution at every wavelet scale~\cite{Addison2017book}, thus turning detection of LAAs and preservation of their morphology easier. Moreover, because EGMs were generated with a sampling rate of 1000~Hz, a five-level decomposition led to the main PLI component of 50~Hz to be mostly located at scale 4 (which covered the frequency range 32.25--62.5 Hz) and its harmonics at scales 3 (62.5--125 Hz) and 2 (125--250~Hz). Then, shrinkage was only required in these scales, since lower frequency information (i.e., scale 5) was undisturbed by the PLI~\cite{Castillo:2013gb}.  As a final step, the denoised signal was reconstructed by applying inverse SWT both to unmodified and shrunk wavelet coefficients for all scales.  
\par
In the literature no systematic guidelines can be found to select an optimal wavelet mother for each application~\cite{Addison2017book}. Although it is often chosen by its similarity to the fundamental pattern to be denoised~\cite{Singh2006}, for a thorough validation of the proposed algorithm all functions from the most common orthogonal wavelet families, including  Daubechies, Biorthogonal, Coiflets, Symlets, Reverse Biorthogonal and Discrete Approximation of Meyer, were tested. Similarly, although well-known alternatives exist to compute the denoising threshold for each wavelet scale in the case of gaussian and white noise~\cite{Bcharri2017}, no many options can be found for suppression of a  sinusoidal interference. Hence, according to a few previous works~\cite{Xu2004,Poornachandra2008}, an adaptive threshold was computed for each level by discarding wavelet coefficients with larger amplitude, such as Figure~\ref{fig:methods}(b) shows for an example. In general terms, for the scale $j$, the threshold $\lambda_j(n)$ was obtained as the result of applying a moving median filtering of 200~ms in length to the absolute value of the wavelet coefficients. This window of 200 ms was selected after several experiments, because it reached the best ability to track amplitude changes in the PLI. 
\par
Finally, to remove low-amplitude wavelet coefficients, a combination of the two most common thresholding functions, i.e., \emph{hard} and \emph{soft} shrinkage approaches~\cite{Donoho1994}, was used to exploit the main characteristics of the EGM morphology. Thus, the main idea was to reduce as much noise as possible from intervals between successive LAAs, but simultaneously preserving these impulsive waveforms. Hence, whereas a hard thresholding function was used in areas associated to the LAAs, a soft shrinkage approach was applied to the remaining regions. To discern in an easy way between both cases, twice the threshold  $\lambda_j(n)$ was experimentally established, such that 
\begin{equation}
\widehat{C_j}(n) = \left\{\begin{array}{ll}
0, &\textrm{if } |\widetilde{C_j}(n)| \leq \lambda_j(n),\\
\widetilde{C_j}(n)-\lambda_j(n), &\textrm{if } \lambda_j(n)<\widetilde{C_j}(n) \leq 2\cdot \lambda_j(n), \\
\widetilde{C_j}(n), &\textrm{if } \widetilde{C_j}(n) > 2\cdot\lambda_j(n),\\
\widetilde{C_j}(n)+\lambda_j(n), &\textrm{if } -2\cdot\lambda_j(n) < \widetilde{C_j}(n) \leq -\lambda_j(n),\\
\widetilde{C_j}(n), &\textrm{if } \widetilde{C_j}(n) < -2\cdot\lambda_j(n),
\end{array} \right.
\end{equation}  
$\widetilde{C_j}(n)$ referring to the set of wavelet coefficients for the scale $j$. The transfer function of this thresholding approach is displayed in Figure~\ref{fig:methods}(c).  

\subsection{Other methods for PLI suppression}
As previously mentioned, EGM recording systems usually include a simple notch filter to reduce the main PLI component of 50~Hz~\cite{Bakker2010}. Clearly, in this way harmonic distortion often contained by the interference cannot be removed. Nonetheless, to recreate this context in a realistic fashion, a similar fixed-bandwidth filter was implemented for comparison with the proposed SWT-based method. Thus, a 2nd-order Butterworth notch filter centered at 50~Hz and with a band-stop of  $\pm$1~Hz was used to remove the PLI from noisy EGM recordings. Despite its limitations, this filtering has been widely applied to many physiological recordings~\cite{Zivanovic2013}. 
\par
As an alternative to this kind of filtering, several adaptive notch filters have been proposed to reduce PLI in the context of the surface ECG. Thus, a nonlinear adaptive filtering was also considered for comparison, given its limited sensitivity to transients~\cite{Sornmo2005}. The method was based on subtracting a sinusoid of 50~Hz, referred to as $v(n)$, to the noisy input signal. To increase the filter effectivity, the amplitude of $v(n)$ was adapted to the PLI through the nonlinear equation 
\begin{equation}
\hat{v}(n) = v(n) + \kappa\cdot\textrm{sgn}(e'(n)), 
\end{equation}
where $\kappa$ controls the filter transient suppression property and was experimentally established to 2~$\mu$V, $\textrm{sgn}(\cdot)$ is the sign function and $e'(n)$ is the first difference of the error between $\tilde{x}(n)$ and $v(n)$~\cite{Sornmo2005}. Finally, the denoised signal was estimated as
\begin{equation}
\hat{x}(n) = \tilde{x}(n) - \hat{v}(n).
\end{equation}
\par
The last method introduced for comparison was a DWT-based algorithm previously proposed to remove PLI in the surface ECG~\cite{Poornachandra2008}. Briefly, the bipolar EGM signal was first decomposed into five wavelet scales and the levels 2, 3 and 4 were shrunk through a hyperbolic thresholding function. In this case, the denoising threshold for each scale was obtained by filtering wavelet coefficients with an $\alpha$-trimmed mean filter (window of 200~ms and $\alpha = 0.3$)~\cite{Poornachandra2008}.

\subsection{Performance assessment}\label{sec:parameters}
The ability of the tested algorithms to preserve the original signal morphology after denoising was analyzed both in time and frequency domains. Thus, common parameters for bipolar EGM characterization were computed from the clean and denoised signals, such as in a previous work~\cite{Martinez2017}. More precisely, the root mean square error (RMSE) between $x(n)$ and  $\hat{x}(n)$ was used to quantify morphological alterations introduced by the denoising algorithm. To obtain a relative measure, this parameter was normalized by the root mean square value of the original signal, i.e.,
\begin{equation}
\textrm{RMSE}=\sqrt{\frac{\sum_{k=1}^N\big(x(k)-\hat{x}(k)\big)^2}{\sum_{k=1}^Nx(k)^2}},
\end{equation}
For the same purpose, an adaptive signed correlation index (ASCI) was also computed from the dichotomization of $x(n)$ and $\hat{x}(n)$~\cite{Lian2010}, such that
\begin{equation}
\textrm{ASCI}\big(x(n),\hat{x}(n)\big)=\frac{1}{N}\sum_{k=1}^{N}x(k)\otimes\hat{x}(k), 
\end{equation}
where the operation $\otimes$ was defined as
\begin{equation}
x(n)\otimes\hat{x}(n)=\left\{
\begin{array}{rl}
1 & \mathrm{if}~|x(n)-\hat{x}(n)| \leq \beta,\\
-1 & \mathrm{if}~|x(n)-\hat{x}(n)| > \beta.
\end{array} \right.
\end{equation}
After several experiments, the optimal value for the threshold $\beta$ was set at 5\% of the standard deviation of $x(n)$. It is interesting to note that, unlike the well-established Pearson's correlation index, the ASCI counts amplitude differences between two signals for their morphological comparison~\cite{Lian2010}.  
\par
On the other hand, given that the EGM fractionation level plays a key role for some protocols of catheter ablation~\cite{Quintanilla2016}, its variation caused by the denoising algorithms was also estimated. For this purpose, the EGM fractionation was quantified by computing its information content through Shannon Entropy (ShEn)~\cite{Ng2010}. In short, the recording was quantized into $L$ uniform and equal-size amplitude levels, so that the repetition rate for the $k$-th step was estimated as its occurrence probability (i.e., $p_k$). Then, the metric was computed as~\cite{Ng2010}
\begin{equation}
\textrm{ShEn}=\frac{-1}{\log{L}}\sum_{k=1}^{L}p_k\cdot\log{p_k}. 
\end{equation}
In this way the metric only obtains a global value for the signal, without paying special attention to the fractionation degree added to the LAAs after denoising. However, these activations contain relevant high-frequency information, which is often altered during EGM preprocessing~\cite{Martinez2017}. Hence, to specifically study how these relevant waveforms were altered, their repetition rate was estimated from the original and denoised signals. Thus, the algorithm proposed by Faes et. al.~\cite{Faes2002} was used to quantify the relative number of similar LAAs in a ensemble of $M$ waves, i.e. 
\begin{equation}
\rho(\xi)=\frac{2}{M\cdot(M-1)}\sum_{i=1}^M\sum_{j=i+1}^M\Theta(\xi-d_{ij}),
\end{equation}
\noindent where $\Theta(x)$ is the Heaviside function, $d_{ij}$ is the distance between the $i$-th and $j$-th activations, computed as the arc cosine of the scalar product of both waveforms, and $\xi$ is the tolerance to decide if they were or not similar. This last value was set at $\pi/3$ radians, such as recommended by the authors~\cite{Faes2002}. Note that in this case no detection of LAAs was required, because their positions were known in the synthesized EGM signals.
\par
From a spectral point of view, the DF has received considerable clinical attention in the last years~\cite{Platonov2014}. Indeed, based on the idea that high-frequency sources can perpetuate the arrhythmia, ablation of atrial areas presenting high DF values has been proposed as a potential treatment for persistent AF~\cite{Gadenz2017}. Hence, in view of its relevant role, the variation experimented by this parameter after denoising was also analyzed. The popular approach introduced by Botteron \& Smith was firstly used to enhance occurrence times of LAAs. Next, the spectral content of the resulting signal was obtained making use of a Welch Periodogram with a Hamming window of 4096 points in length, a 50\% overlapping between adjacent windowed sections and a 10240-point FFT~\cite{Everett2001}. The DF was then estimated as the frequency presenting the highest energy. Finally, the variation observed after denoising in the power spectral density ratio between the area under the DF together with its three first harmonics (computed over a $\pm$1 Hz window) and the total area from 2.5~Hz to the fifth harmonic peak, was also computed~\cite{Everett2001}. The rationale for using this metric, named organization index (OI), relies on the fact that it has been vastly used in clinical studies as a marker of AF organization~\cite{Baumert2016}.
\par
Finally, note that statistically significant differences between the values of ShEn, $\rho$, DF and OI computed both from the original and denoised signals were estimated by using a parametric Student's $t$-test. According to Shapiro-Wilk and Levene tests, data were normal and homoscedastic for most cases. 

\section{Results}\label{sec:results}
\subsection{PLI effect on the clean EGM signals}
\begin{figure*}[tb!]
\flushright
\includegraphics[width=0.6\textwidth,keepaspectratio]{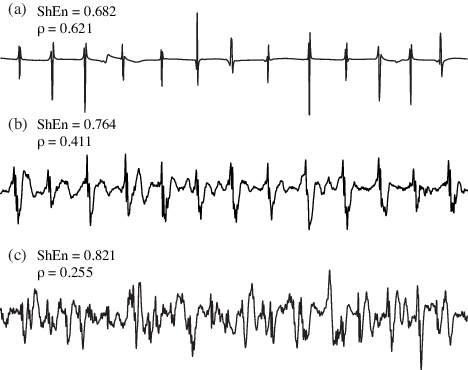}
\caption{Three pseudo-real EGM signals synthesized with different levels of organization and fractionation through the method described in Section~\S\ref{sec:gen_EGM}. The values of ShEn and $\rho$ computed for each signal are also displayed.}\label{fig:ex_EGM}
\end{figure*}

For a thorough validation of the tested denoising algorithms, clean EGMs were synthesized as realistic as possible. Indeed, a broad variety of representative morphologies were generated, thus collecting a balanced set of organized and fractionated EGM recordings. In terms of ShEn and $\rho$, these signals reported values between 0.889 and 0.624, with a median of 0.789, and between 0.812 and 0.123, with a median of 0.488, respectively. Regarding the OI, a wide range of values from 0.191 to 0.564 (with median of 0.356) were also noticed. In the three cases, these values were closely similar to those reported by other authors for real EGMs~\cite{Ng2010,Faes2002,Everett2001}. Examples of three obtained clean EGMs with different levels of organization are displayed in Figure~\ref{fig:ex_EGM}.   

%%%%%%%%%%%%%%%%%%%
%%%%%%%%%%%%%%%%%%%

On the other hand, the PLI effect on these EGM recordings is presented in Table~\ref{tab:noise}, where the symbol $\Delta$ refers to the relative variation between original and noisy signals. As expected, for all tested scenarios of PLI the lower the SIR, the higher the alteration reported by the assessment parameters. In fact, whereas negligible changes lower than 5\% were only noticed for SIR values larger than 25~dB, more pronounced errors were observed for higher levels of noise. To this respect, it is worth noting that moderate levels of SIR, between 15 and 20~dB, were sufficient to provide statistical differences between values of ShEn, $\rho$ and OI computed from the clean and noisy signals in every scenario of PLI. Moreover, excluding the DF, deviations between 10 and 35\% and between 20 and 65\% were seen in the remaining assessment parameters for SIR values of 10 and 5~dB, respectively.   

%%%%%%%%%%%%%%%%%%% 

\begin{table*}[h!]
\caption{Relative variations observed in the assessment parameters when they were computed from the clean and noisy EGM recordings. One hundred 10 s-length EGMs were analyzed. Statistically significant differences ($p<0.05$) between the values of ShEn, $\rho$, DF and OI computed from the clean and noisy signals are marked by the symbol $^{\dagger}$.}\label{tab:noise}%
\vspace{17.5cm}\hspace*{10.2cm}
%\begin{flushright}
\begin{rotate}{90}
\scriptsize%\centering
\begin{tabular}{lccccccc}
\hline
\hline
\textbf{Scenario} & \textbf{SIR (dB)} & \textbf{RMSE (\%)} & \textbf{ASCI (\%)} & \textbf{$\Delta$ShEn (\%)} & \textbf{$\Delta\mathbf{\rho}$ (\%)} &  \textbf{$\Delta$DF (\%)} &\textbf{$\Delta$OI (\%)} \\
\hline
 \multirow{6}{*}{\begin{minipage}{2.5cm}Common\\ PLI\end{minipage}} &
30   & 0.10 $\pm$ 4.30$\times10^{-5}$ & 100 $\pm$ 0.00  & 1.22 $\pm$ 1.39  & 2.49 $\pm$ 2.51 & 0.16 $\pm$ 0.86    & 3.24 $\pm$ 0.98\\
& 25   & 0.32 $\pm$ 1.44$\times10^{-4}$ & 100 $\pm$ 0.04 & 3.15 $\pm$  2.91    & 4.99 $\pm$ 11.97 & 0.16 $\pm$ 0.86    & 5.90 $\pm$ 1.67$^{\dagger}$\\
& 20   & 1.01 $\pm$ 4.40$\times10^{-4}$ & 99.97 $\pm$ 2.28   & 7.08 $\pm$ 5.59$^{\dagger}$   & 14.46 $\pm$ 19.11 & 0.18 $\pm$ 2.50 & 10.92 $\pm$ 2.90$^{\dagger}$\\
& 15   & 3.16 $\pm$ 1.44$\times10^{-3}$ & 81.38 $\pm$ 8.07 & 13.79 $\pm$ 9.40$^{\dagger}$   & 23.21 $\pm$ 12.95$^{\dagger}$  & 6.72 $\pm$ 23.22  & 20.48 $\pm$ 4.57$^{\dagger}$ \\
& 10   & 10.01 $\pm$ 4.53$\times10^{-3}$ & 39.78 $\pm$ 8.74   & 23.28 $\pm$ 13.99$^{\dagger}$  & 27.66 $\pm$ 12.77$^{\dagger}$ & 7.12 $\pm$ 21.69  & 37.34 $\pm$ 5.99$^{\dagger}$\\
& 5 & 31.63 $\pm$ 1.0$\times10^{-2}$ & 23.04 $\pm$ 5.96 & 35.39 $\pm$ 18.00$^{\dagger}$ & 22.49 $\pm$ 14.21$^{\dagger}$ & 7.08 $\pm$ 20.63 & 60.47 $\pm$ 6.15$^{\dagger}$\\
\hline
 \multirow{6}{*}{\begin{minipage}{2.5cm}PLI with time-\\varying amplitude\end{minipage}}
& 30 & 0.10 $\pm$ 4.22$\times10^{-5}$  & 100.00 $\pm$ 0.00 & 1.17 $\pm$ 1.48 & 2.02 $\pm$ 2.77 & 0.74 $\pm$ 4.21 & 2.40 $\pm$ 1.04 \\
& 25 & 0.32 $\pm$ 1.34$\times10^{-4}$  & 100.00 $\pm$ 0.00 & 2.91 $\pm$ 2.95 & 11.60 $\pm$ 16.98 & 0.70 $\pm$ 4.21 & 4.54 $\pm$ 1.43$^{\dagger}$ \\
& 20 & 1.00 $\pm$ 4.22$\times10^{-4}$  & 96.56 $\pm$ 3.72 & 6.63 $\pm$ 5.50$^{\dagger}$ & 19.69 $\pm$ 18.14$^{\dagger}$ & 0.85 $\pm$ 4.45 & 8.45 $\pm$ 3.07$^{\dagger}$ \\
& 15 & 3.16 $\pm$ 1.32$\times10^{-3}$  & 81.18 $\pm$ 5.25 & 13.23 $\pm$ 8.98$^{\dagger}$ & 29.07 $\pm$ 11.72$^{\dagger}$ & 1.87 $\pm$ 8.13 & 16.16 $\pm$ 4.66$^{\dagger}$ \\
& 10 & 10.00 $\pm$ 4.09$\times10^{-3}$  & 65.24 $\pm$ 5.54 & 22.68 $\pm$ 12.85$^{\dagger}$ & 26.02 $\pm$ 12.54$^{\dagger}$ & 5.77 $\pm$ 17.00 & 29.94 $\pm$ 5.91$^{\dagger}$ \\
& 5 & 31.62 $\pm$ 1.34$\times10^{-2}$  & 50.23 $\pm$ 3.47 & 34.72 $\pm$ 16.28$^{\dagger}$ & 21.54 $\pm$ 13.12$^{\dagger}$ & 3.69 $\pm$ 10.42 & 51.56 $\pm$ 6.31$^{\dagger}$ \\
\hline
 \multirow{6}{*}{\begin{minipage}{2.5cm}PLI with \\frequency deviation\end{minipage}} 
 & 30 & 0.10 $\pm$ 4.25$\times10^{-5}$  & 100.00 $\pm$ 0.00 & 1.25 $\pm$ 1.48 & 1.56 $\pm$ 1.89 & 0.16 $\pm$ 0.86 & 3.31 $\pm$ 1.08 \\
& 25 & 0.32 $\pm$ 1.29$\times10^{-4}$  & 100.00 $\pm$ 0.00 & 3.14 $\pm$ 2.97 & 5.71 $\pm$ 9.81 & 0.74 $\pm$ 4.21 & 5.93 $\pm$ 1.72$^{\dagger}$ \\
& 20 & 1.00 $\pm$ 4.25$\times10^{-4}$  & 99.98 $\pm$ 0.12 & 7.09 $\pm$ 5.53$^{\dagger}$ & 13.42 $\pm$ 16.35 & 2.94 $\pm$ 19.22 & 11.48 $\pm$ 2.74$^{\dagger}$ \\
& 15 & 3.16 $\pm$ 1.36$\times10^{-3}$  & 81.53 $\pm$ 17.11 & 13.78 $\pm$ 9.05$^{\dagger}$ & 21.34 $\pm$ 17.43$^{\dagger}$ & 7.62 $\pm$ 24.39 & 20.50 $\pm$ 4.43$^{\dagger}$ \\
& 10 & 10.00 $\pm$ 4.16$\times10^{-3}$  & 40.00 $\pm$ 8.27 & 23.55 $\pm$ 12.79$^{\dagger}$ & 27.01 $\pm$ 12.18$^{\dagger}$ & 6.78 $\pm$ 23.80 & 37.20 $\pm$ 5.71$^{\dagger}$ \\
& 5 & 31.63 $\pm$ 1.37$\times10^{-2}$  & 22.69 $\pm$ 4.26 & 35.45 $\pm$ 16.51$^{\dagger}$ & 22.47 $\pm$ 13.16$^{\dagger}$ & 9.51 $\pm$ 27.32 & 60.55 $\pm$ 6.57$^{\dagger}$ \\
\hline
 \multirow{6}{*}{\begin{minipage}{2.5cm}PLI with \\strong harmonics\end{minipage}} 
 & 30 & 0.10 $\pm$ 4.60$\times10^{-5}$  & 100.00 $\pm$ 0.00 & 1.23 $\pm$ 1.43 & 2.55 $\pm$ 3.36 & 0.16 $\pm$ 0.86 & 2.88 $\pm$ 0.93 \\
& 25 & 0.32 $\pm$ 1.46$\times10^{-4}$  & 100.00 $\pm$ 0.01 & 3.11 $\pm$ 2.92 & 4.70 $\pm$ 6.88 & 0.16 $\pm$ 0.86 & 5.29 $\pm$ 1.54$^{\dagger}$ \\
& 20 & 1.00 $\pm$ 4.49$\times10^{-4}$  & 98.17 $\pm$ 2.71 & 7.06 $\pm$ 5.66$^{\dagger}$ & 13.11 $\pm$ 17.06 & 2.95 $\pm$ 19.22 & 9.77 $\pm$ 2.68$^{\dagger}$ \\
& 15 & 3.16 $\pm$ 1.44$\times10^{-3}$  & 82.76 $\pm$ 7.85 & 13.95 $\pm$ 9.38$^{\dagger}$ & 29.68 $\pm$ 17.18$^{\dagger}$ & 5.88 $\pm$ 23.65 & 18.53 $\pm$ 4.65$^{\dagger}$ \\
& 10 & 10.00 $\pm$ 3.89$\times10^{-3}$  & 56.62 $\pm$ 9.60 & 24.02 $\pm$ 13.71$^{\dagger}$ & 39.92 $\pm$ 12.44$^{\dagger}$ & 5.86 $\pm$ 23.65 & 33.00 $\pm$ 5.42$^{\dagger}$ \\
& 5 & 31.63 $\pm$ 1.37$\times10^{-2}$  & 33.46 $\pm$ 7.55 & 37.28 $\pm$ 18.01$^{\dagger}$ & 39.91 $\pm$ 17.93$^{\dagger}$ & 8.17 $\pm$ 28.24 & 55.34 $\pm$ 6.26$^{\dagger}$ \\
\hline
 \multirow{6}{*}{\begin{minipage}{2.5cm}Real PLI\end{minipage}} 
& 30 & 0.10 $\pm$ 4.40$\times10^{-5}$  & 100.00 $\pm$ 0.00 & 1.24 $\pm$ 1.37 & 2.38 $\pm$ 3.07 & 0.12 $\pm$ 0.82 & 3.31 $\pm$ 1.03 \\
& 25 & 0.32 $\pm$ 1.30$\times10^{-4}$  & 99.99 $\pm$ 0.05 & 3.12 $\pm$ 2.93 & 7.72 $\pm$ 13.80 & 0.16 $\pm$ 0.86 & 6.09 $\pm$ 1.75$^{\dagger}$ \\
& 20 & 1.00 $\pm$ 4.37$\times10^{-4}$  & 98.21 $\pm$ 2.20 & 7.05 $\pm$ 5.62$^{\dagger}$ & 22.45 $\pm$ 19.24$^{\dagger}$ & 0.26 $\pm$ 0.95 & 11.44 $\pm$ 3.16$^{\dagger}$ \\
& 15 & 3.16 $\pm$ 1.39$\times10^{-3}$  & 82.33 $\pm$ 8.06 & 14.12 $\pm$ 9.54$^{\dagger}$ & 34.20 $\pm$ 13.31$^{\dagger}$ & 0.83 $\pm$ 3.49 & 21.50 $\pm$ 4.40$^{\dagger}$ \\
& 10 & 10.00 $\pm$ 4.43$\times10^{+3}$  & 55.41 $\pm$ 8.76 & 24.43 $\pm$ 13.75$^{\dagger}$ & 39.25 $\pm$ 10.59$^{\dagger}$ & 8.29 $\pm$ 25.08 & 38.13 $\pm$ 5.56$^{\dagger}$ \\
& 5 & 31.62 $\pm$ 1.45$\times10^{-2}$  & 33.15 $\pm$ 6.08 & 37.54 $\pm$ 17.91$^{\dagger}$ & 35.89 $\pm$ 17.33$^{\dagger}$ & 6.01 $\pm$ 18.90 & 61.73 $\pm$ 5.33$^{\dagger}$ \\
\hline
\hline
\end{tabular}
\end{rotate}
%\end{flushright}
\end{table*}
 %%%%%%%%%%%%%%%%%%%
 
\subsection{Denoising of a common PLI}
\begin{figure*}[tb!]
\centering
\includegraphics[width=\textwidth,keepaspectratio]{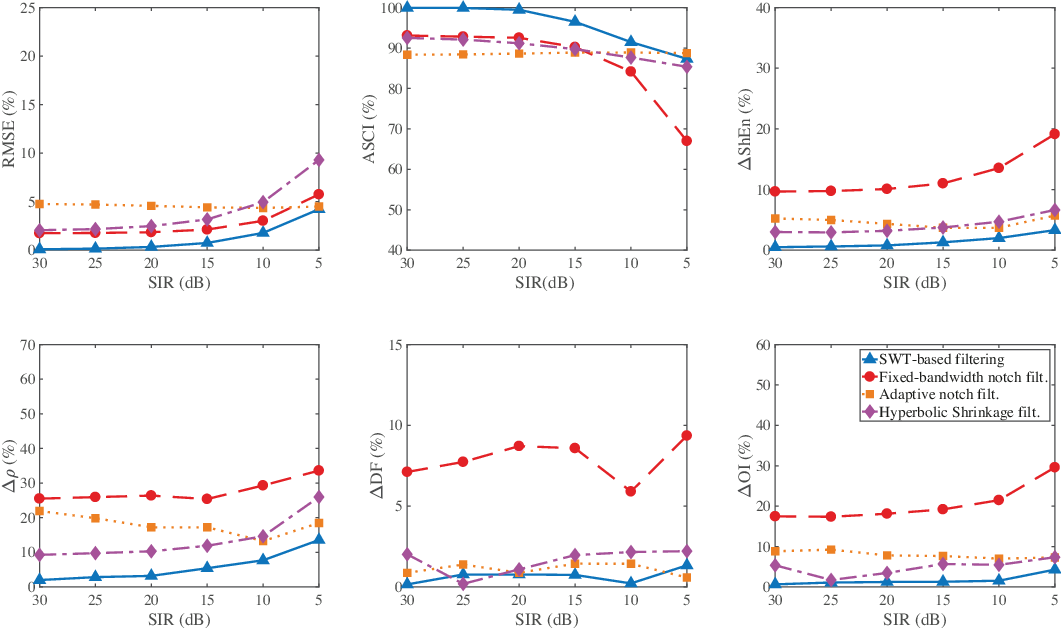}
\caption{Rate and morphological variations provoked by the analyzed methods after denoising of a common PLI. Different levels of noise (SIR) were considered and the original and denoised signals were compared in terms of the parameters RMSE, ASCI, ShEn, $\rho$, DF and OI. Note that average values for the analyzed EGM recordings are displayed.}   
\label{fig:results_common}
\end{figure*}

When all wavelet functions mentioned in Section~\S\ref{sec:shrinking} were tested as mother wavelet for the proposed SWT-based algorithm, only negligible differences were noticed among the obtained results. Thus, only the best outcomes, obtained with a second-order Coiflet function, are displayed in Figure~\ref{fig:results_common}. As can be seen, the algorithm was able to reduce very high levels of PLI, without altering substantially the original EGM morphology. In fact, even for a SIR of 5~dB, variations lower than 5\% for RMSE and ShEn and than 13\% for ASCI and $\rho$ were respectively noticed. From a spectral point of view, the proposed SWT-based algorithm also proved to preserve mostly the original EGM frequency content. Thus, both DF and OI reported changes lower than 3\% regardless of the considered SIR value. Moreover, it is interesting to note that no statistically significant differences were noticed in terms of ShEn, $\rho$, DF and OI for any level of noise. 

%%%%%%%%%%%%%%%%%%%
%%%%%%%%%%%%%%%%%%%

Additionally, Figure~\ref{fig:results_common} also shows that the described results were better than those obtained by the DWT-based hyperbolic thresholding, as well as by the fixed-bandwidth and adaptive notch filters. Indeed, the hyperbolic shrinkage results as a function of SIR were quite similar to the proposed SWT-based algorithm, but with larger variations about 2.5--5\%, 3--13\%, 2--3.5\%, 8--10\% 1--2\% and 1.5--5\% for the parameters RMSE, ASCI, ShEn, $\rho$, DF and OI, respectively. Regarding the adaptive filtering, its behavior was stable for every SIR value, thus showing a worse relative performance, with respect to the SWT-based method, for moderate and low levels of PLI. Although this filtering reported slight variations in most assessment parameters, it was featured by adding notable fractionation in LAAs. Thus, statistically significant variations in the parameter $\rho$, about 20\%, were noticed between the original and denoised EGM recordings. Finally, whereas the fixed-bandwidth filtering reported discrete variations in terms of RMSE and ASCI, its performance was fairly poorer in terms of the remaining assessment parameters. In fact, the indices ShEn, $\rho$ and OI reported statistically significant differences for every level of noise. For the DF, low changes between 5 and 8\% were noticed, thus highlighting that this parameter was the less sensitive to the PLI. Moreover, the observed variations were due to the misidentification of the first harmonic as the DF for a limited number of EGM recordings, i.e. 0, 6, 4, 6 and 16\% for SIR values of 30, 25, 20, 15, 10 and 5~dB, respectively.

\subsection{Denoising of more complex scenarios of PLI}
\begin{figure*}[tbp!]
\centering
\includegraphics[width=\textwidth,keepaspectratio]{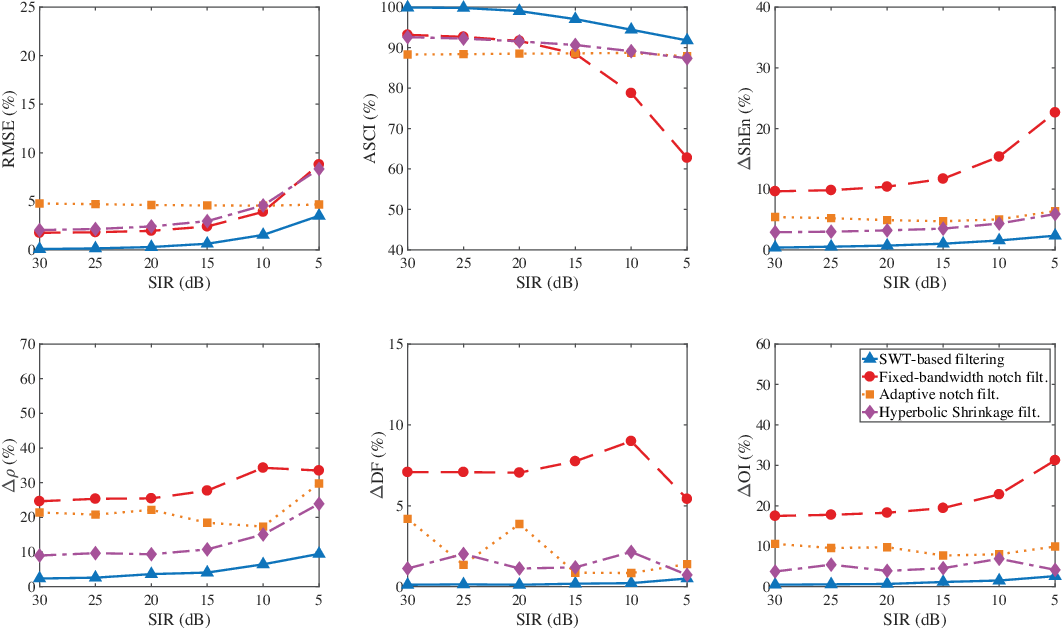}
\caption{Rate and morphological variations provoked by the analyzed methods after denoising of a PLI with time-varying amplitude. Different levels of noise (SIR) were considered and the original and denoised signals were compared in terms of the parameters RMSE, ASCI, ShEn, $\rho$, DF and OI. Note that average values for the analyzed EGM recordings are displayed. }\label{fig:results_ampPLI}
\end{figure*}

\begin{figure*}[tbp!]
\centering
\includegraphics[width=\textwidth,keepaspectratio]{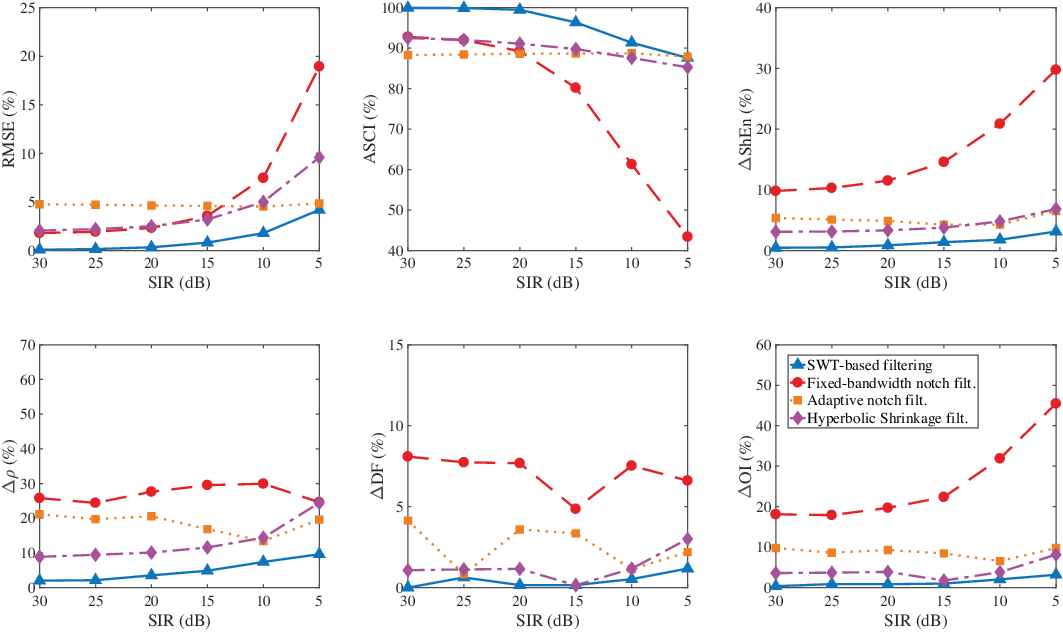}
\caption{Rate and morphological variations provoked by the analyzed methods after denoising of a PLI with frequency deviations between $-$3 and 3~Hz of its main component and harmonics. Different levels of noise (SIR) were considered and the original and denoised signals were compared in terms of the parameters RMSE, ASCI, ShEn, $\rho$, DF and OI. Note that average values for the analyzed EGM recordings are displayed.} \label{fig:results_desvfrecPLI}
\end{figure*}

For any tested additional scenario of PLI the proposed SWT-based algorithm presented a similar behavior, compared with the common interference. Contrarily, the performance of the DWT-based hyperbolic thresholding, as well as of the fixed-bandwidth and adaptive notch filters, was considerably degraded under some conditions. To this respect, Figure~\ref{fig:results_ampPLI} displays how the denoising algorithms responded when the PLI showed sudden and sinusoidal amplitude variations. Apart from the fact that fixed-bandwidth notch filtering introduced slightly variations in all metrics for SIR values of 5 and 10~dB, no significant differences are seen in comparison with Figure~\ref{fig:results_common}. In a similar line, Figure~\ref{fig:results_desvfrecPLI} shows a notably worsened performance for the fixed-bandwidth notch filter when maximum frequency deviations of $\pm$ 3~Hz were considered for the main component of 50~Hz and its harmonics. Indeed, the parameters RMSE, ASCI and OI reported variations between 15 and 25\% regarding the common PLI analysis. Moreover, statistically significant differences were also noticed between the values of ShEn, $\rho$ and OI computed between the original and denoised signals with this kind of filtering. On the contrary, the proposed SWT-based algorithm, the hyperbolic thresholding and the adaptive filtering reported the same behavior in the described context as for the common PLI.     

%%%%%%%%%%%%%%%%%%%
%%%%%%%%%%%%%%%%%%%

%%%%%%%%%%%%%%%%%%%
%%%%%%%%%%%%%%%%%%%

\begin{figure*}[tbp!]
\centering
\includegraphics[width=\textwidth,keepaspectratio]{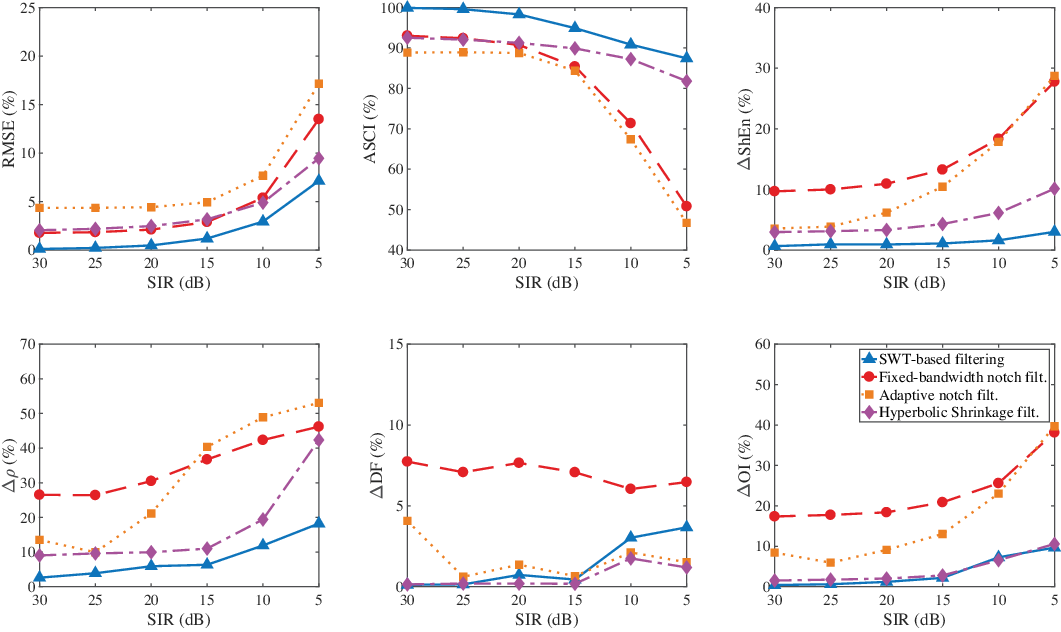}
\caption{Rate and morphological variations provoked by the analyzed methods after denoising of a PLI with increased harmonic content between 5 and 10 times the recommended by the standard EN-50160~\cite{Cenelec:1999tn}. Different levels of noise (SIR) were considered and the original and denoised signals were compared in terms of the parameters RMSE, ASCI, ShEn, $\rho$, DF and OI. Note that average values for the analyzed EGM recordings are displayed.} \label{fig:results_harmPLI}
\end{figure*}

\begin{figure*}[tbp!]
\centering
\includegraphics[width=\textwidth,keepaspectratio]{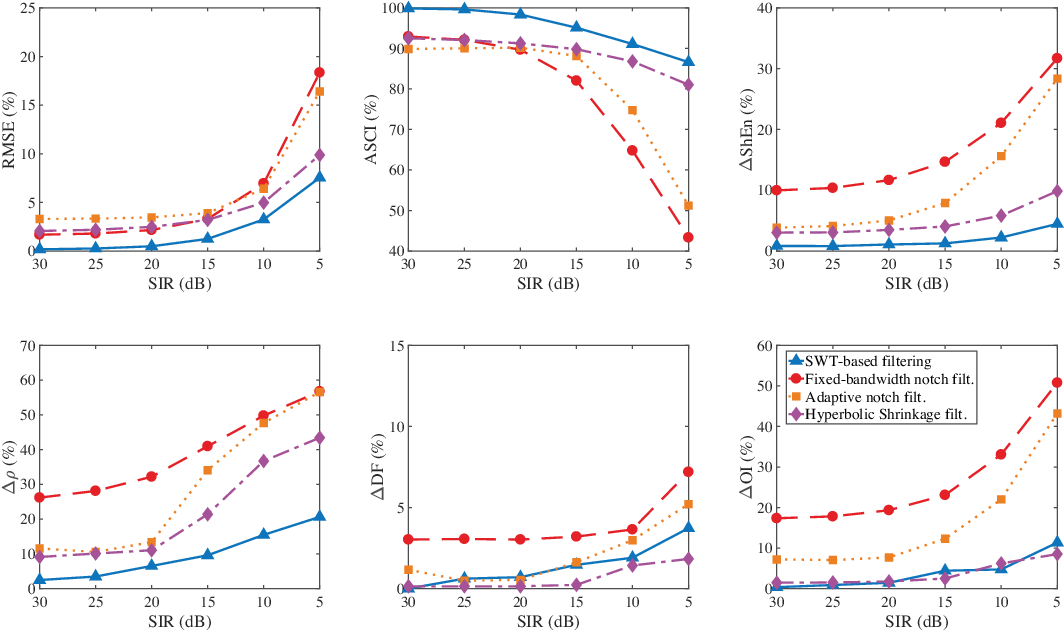}
\caption{Rate and morphological variations provoked by the analyzed methods after denoising of a real PLI. Different levels of noise (SIR) were considered and the original and denoised signals were compared in terms of the parameters RMSE, ASCI, ShEn, $\rho$, DF and OI. Note that average values for the analyzed EGM recordings are displayed.} \label{fig:results_realPLI}
\end{figure*}

On the other hand, when the PLI harmonic distortion was notably increased regarding the limit established by the standard EN-50160, the SWT-based algorithm performance still remained almost unchanged, such as Figure~\ref{fig:results_harmPLI} displays. Thus, only variations about 5\% were noticed in the parameters RMSE, $\rho$ and OI for the noisiest case with SIR of 5~dB. Similarly, slight changes were also observed for the DWT-based hyperbolic thresholding in all assessment parameters, apart from the index $\rho$. In this case, statistically significant increases about 7 and 15\% were seen for  SIR values of 10 and 5~dB, respectively. Contrarily, the obtained results by both fixed-bandwidth and adaptive notch filters were notably poorer, compared with the common PLI, especially for the highest levels of noise. Indeed, in both cases the metric ASCI reported statistically significant variations higher than 25\% and the remaining parameters between 5 and 30\% for a SIR value of 5~dB. Finally, it is interesting to note that a very similar performance was also provided by the denoising algorithms when a real PLI was considered. Precisely, no significant differences can be found between Figures~\ref{fig:results_harmPLI} and~\ref{fig:results_realPLI}. Moreover, both simulated ($-4.52 \pm 2.21$~dB) and real ($-5.15 \pm 1.98$~dB) interferences contained comparable levels of average harmonic distortion.    

%%%%%%%%%%%%%%%%%%%
%%%%%%%%%%%%%%%%%%%

%%%%%%%%%%%%%%%%%%%
%%%%%%%%%%%%%%%%%%%

\begin{figure*}[tbp!]
\centering
\includegraphics[width=\textwidth,keepaspectratio]{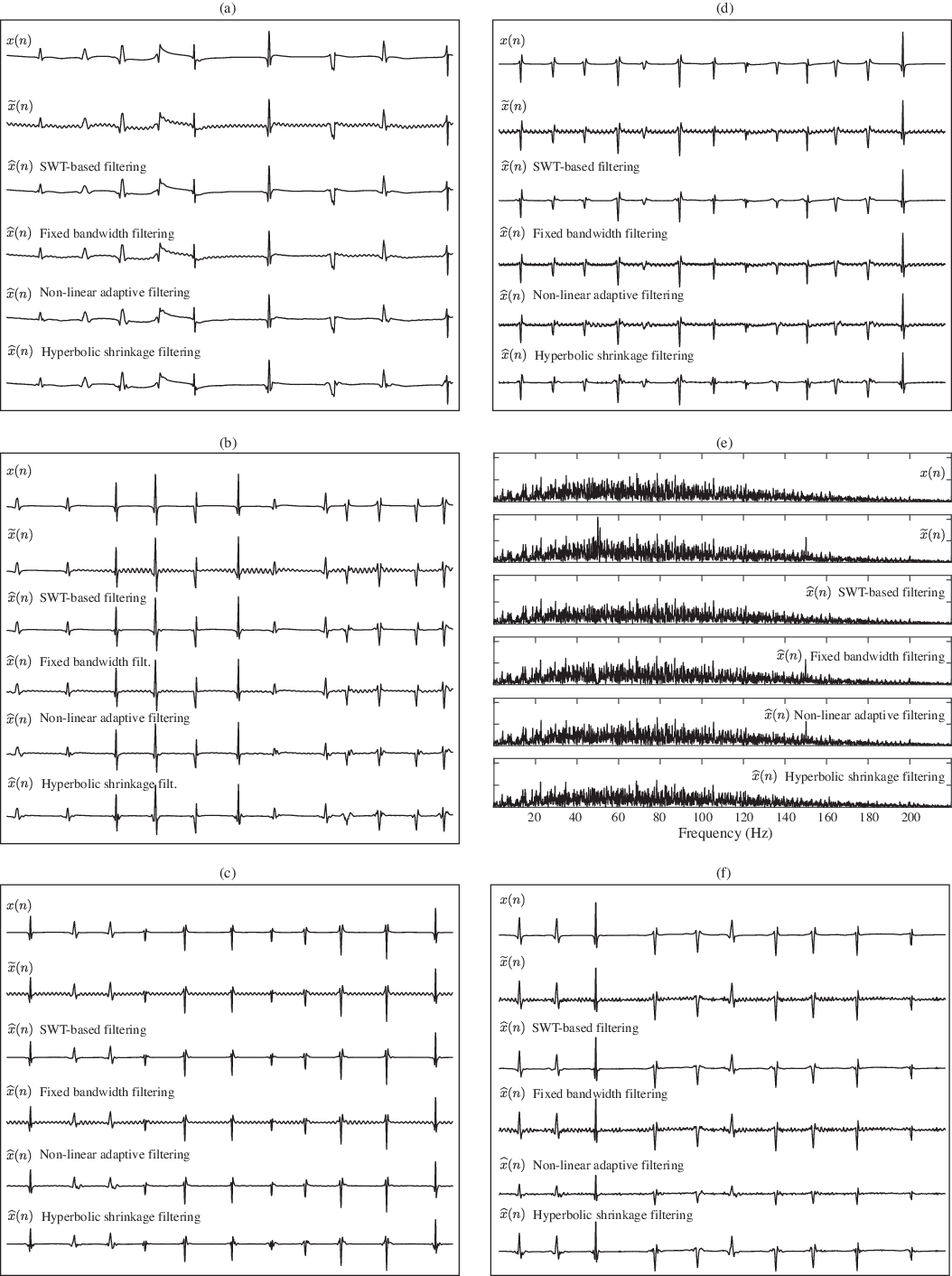}
\caption{Typical examples displaying the four methods' performance for all the considered interferences: (a) Common PLI; (b) PLI with time-varying amplitude; (c) PLI with a maximum frequency deviation of $\pm$3~Hz; (d) PLI with a harmonic distortion higher than the recommended by the standard EN-50160; (e) Spectral distributions for all the signals presented in (d); and (f) real PLI. Note that 2 second-length EGM recordings with an initial SIR value of 10 dB are always presented.}\label{fig:example}
\end{figure*}

As a graphical summary, Figure~\ref{fig:example} shows the signals obtained with the denosing algorithms from typical EGMs corrupted with all the tested interferences for a SIR value of 10 dB. Note that, for illustrative purposes, spectral distributions from the denoised signals are also presented in the case of a PLI with harmonics stronger than the common interference.  
 
%%%%%%%%%%%%%%%%%%% 
%%%%%%%%%%%%%%%%%%%

\subsection{Denoising of real EGM recordings}
\begin{figure*}[tbp!]
\centering
\includegraphics[width=\textwidth,keepaspectratio]{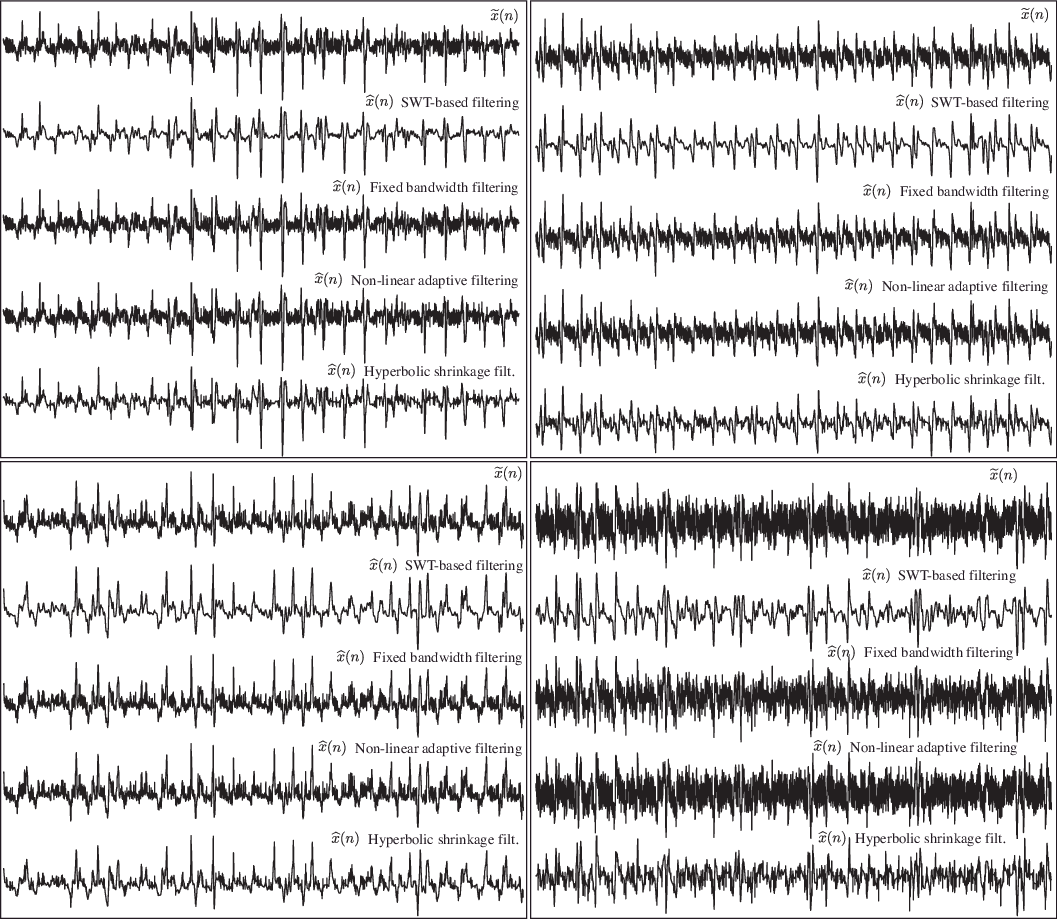}
\caption{Examples displaying the four methods' performance over real EGM excerpts extracted from the freely available IAFDB~\cite{Goldberger2000}. As can be seen, the denoised signal with the proposed SWT-based method clearly contains less residual PLI than those resulting from the remaining algorithms.}\label{fig:real}
\end{figure*}

Given that cardiac electrical activity and PLI are unavoidably recorded together in real EGMs, all the parameters that have been previously used to assess denoising performance over synthesized EGMs cannot be employed in this case. Hence, only visual validation of the results provided by denoising methods is possible. To this respect, Figure~\ref{fig:real} shows several representative examples for real EGM signals. As can be seen, regardless of the EGM morphology, fractionation degree and noise level, the denoised signal with the proposed SWT-based method clearly contains less residual PLI than those resulting from the other algorithms. Moreover, as in the analysis of synthesized EGM signals, fixed-bandwidth and adaptive notch filtering approaches were unable to remove most of the noise.

%%%%%%%%%%%%%%%%%%%
%%%%%%%%%%%%%%%%%%%

\section{Discussion}\label{sec:discussion} 

\subsection{PLI effect on the clean EGM signals}
To a lesser extent than for unipolar EGMs and other surface physiological signals, the PLI also disturbs bipolar EGM acquisition~\cite{Venkatachalam2011a}. Moreover, the obtained results suggest that a little interference is large enough to conceal the accurate characterization and interpretation of these recordings. In fact, Table~\ref{tab:noise} shows that moderate SIR values about 15--20~dB are sufficient to provoke statistically significant variations in the parameters ShEn, $\rho$ and OI for every scenario of PLI. This finding agrees with previous works which have recommended a signal-to-noise ratio of 20~dB as the lowest threshold to discern small changes in bipolar EGM recordings~\cite{Venkatachalam2011a}. 
\par
Additionally, the outcomes displayed in Table~\ref{tab:noise} also highlight the need of suitably  preprocessing bipolar EGM signals to successfully guide catheter ablation. To this respect, ablations based on targeting complex and fractionated EGMs (CFAEs)~\cite{Nademanee2004,Ganesan2013a}, areas showing high-frequency activity~\cite{Atienza2009} or regions with different features of organization~\cite{Jarman2014}, and rate and wave-similarity~\cite{Ravelli2014abl} have been proposed as additional ablation alternatives, beyond pulmonary veins isolation, for patients with persistent AF. Recently, some authors have also suggested that low-voltage areas could play a key role in the maintenance of the arrhythmia~\cite{Kottkamp2016}. Thus, Honarbakhsh et al.~\cite{Honarbakhsh2018} have proven that rate-dependent conduction velocity slowing sites are predominantly confined to low-voltage regions (0.2--0.5 mV), and the resultant conduction velocity heterogeneity may promote formation of drivers in AF. However, all these ablation strategies could identify mistakenly inadequate critical atrial regions that supposedly are driving AF. This is due to their blind trust on EGMs that would be altered by the presence of PLI because they are processed with conventional cardiac signal acquisition equipments. As a consequence, the validation of these ablation protocols would be clinically questionable.

\subsection{Main findings about the regular filtering}
According to previous works~\cite{Jadidi2013,Venkatachalam2011a}, the obtained results suggest that the fixed-bandwidth Butterworth notch filter substantially disturbs the original EGM morphology. Thus, alterations higher than 20\% were observed for the parameters ASCI, ShEn, $\rho$ and OI, when noisy recordings with reduced SIR values of 10 and 5~dB were considered (see Figure~\ref{fig:results_common}). This poor outcome can be easily explained by the massive removal of EGM information about 50~Hz. Moreover, this kind of filtering also introduces ringing in response to steep and quick components of the input signal, such as the LAAs. Regrettably, this perturbation remains even with the absence of PLI~\cite{Hamilton1996,Warmerdam2017}, thus modifying substantially the original EGM morphology for every value of SIR. This aspect is reflected in Figure~\ref{fig:results_common} by high variations in the metric $\rho$. Another relevant limitation of this denoising method is its inability to reduce successfully the PLI under frequency deviations in the range of $\pm$~3~Hz. In this case, its behavior was considerably worsened, such as Figure~\ref{fig:results_desvfrecPLI} displays. 
\par
This poor performance was improved by the nonlinear adaptive filtering for most scenarios of PLI. Thus, slight variations in the parameters RSME and ASCI, and no statistically significant differences between the values of ShEn, DF and OI computed from the original and denoised signals were noticed for the common PLI (see Figure~\ref{fig:results_common}), as well as for interferences with time-varying amplitudes and frequency deviations (see Figs.~\ref{fig:results_ampPLI} and~\ref{fig:results_desvfrecPLI}). Nonetheless, in all cases the method significantly altered the original morphology of the LAAs, because variations about 20\% (o higher) were observed for the parameter $\rho$. This outcome could be mainly explained by the interference of the LAAs in the adaptation procedure, because high learning rates are required to track sudden changes in the PLI~\cite{Martens2006,Warmerdam2017}. Although a better result has been obtained by reducing the learning rate during QRS complexes in the ECG signal~\cite{Martens2006}, identification of LAAs from the EGM is often a non-trivial task~\cite{Chang2011,Ng2014}. Other solution to reduce the QRS interference in the adaptation procedure is the use of Kalman filtering. Several variants can be found in the literature for the PLI suppression from the ECG signal~\cite{Warmerdam2017}. However, for a successful performance of this methodology, observational noise must mandatorily be white. This noise represents all non-PLI signals, including physiological information carried by the recording, and thereby an aggressive low-pass filtering below 30~Hz has been used for ECG whitening~\cite{Warmerdam2017}. Unfortunately, this approach is not applicable to the EGM signal, because it contains relevant spectral information for frequencies above 30~Hz~\cite{Venkatachalam2011a}. Hence, other alternatives need to be studied in the future.

\subsection{Main findings about the proposed SWT-based algorithm}
The limitations presented both by  fixed-bandwidth and adaptive filtering approaches are mostly overcome by the proposed methodology based on SWT shrinkage. In fact, the algorithm has reported a notably better performance than both filters for every scenario of PLI (see Figs.~\ref{fig:results_common}--\ref{fig:results_realPLI}). More precisely, variations lower than 12\% were observed for all the assessment parameters and  most of the tested situations, only detecting alterations about 20\% for the metric $\rho$ when the PLI presented significantly strong harmonics (see Figure~\ref{fig:results_harmPLI}). Furthermore, the method has also outperformed the other analyzed wavelet-based denoising approach, i.e. the DWT-based hyperbolic thresholding. This outcome could be explained by the fact that the SWT-based technique includes some improvements for an enhanced trade-off between noise reduction and original EGM morphology preservation. On the one hand, the SWT is used instead of DWT, thus discarding decimation for all the wavelet scales. The number of coefficients associated with sharp transitions in the LAAs is unaltered for every wavelet level and, therefore, morphology of these relevant waveforms is less disturbed by the shrinkage approach than in the case of the DWT. On the other hand, instead of the hyperbolic thresholding function, a novel combination of soft and hard approaches has been proposed for improved morphology preservation of sudden transitions in the LAAs (see Figure~\ref{fig:methods}(c)). It should be noted that the hyperbolic thresholding function is mainly featured by introducing a smooth transition between zeroed and preserved wavelet coefficients, thus losing the ability of hard thresholding to retain high-frequency information intact~\cite{Poornachandra2008}. 
\par
The proposed SWT-based algorithm is also characterized by its ability to remove the main PLI component of 50~Hz and its harmonic content in a single step. In fact, its behavior is mainly unaltered in the presence or absence of harmonics (compare Figs.~\ref{fig:results_common} and~\ref{fig:results_harmPLI}), as well as when any frequency component (50~Hz or its harmonics) presents large deviations (compare Figs.~\ref{fig:results_common} and~\ref{fig:results_desvfrecPLI}). This feature represents a worthwhile advantage regarding common approaches based on filtering. Although in this study removal of the main PLI component through regular filtering has only been analyzed, cascading fixed-bandwidth or adaptive notch filters could have been used to suppress the remaining harmonic distortion~\cite{Allen2009,Costa2009}. However, in this case a filtering structure is required for each spectral component that has to be removed, thus making it impossible to reduce those interharmonics without \emph{a priori} knowledge about their location. Another limitation of this approach is the impossibility to connect a high number of filters in cascade, since its complexity and computational load would exponentially increase, especially for adaptive filtering where several parameters have to be estimated for each spectral component. 

\subsection{Comparison of the proposed method with its predecessor}
A first version of the proposed SWT-based algorithm was previously introduced in~\cite{Martinez2017-cinc}. However, the method presented the important limitation of requiring proper detection of all LAAs to preserve their morphology. Indeed, the undetected activations were considered as noise and, then, removed. Thus, the original EGM morphology could be notably altered when dealing with CFAEs because, as previously mentioned, detection of LAAs is a complex task  in this case~\cite{Chang2011,Ng2014}. To overcome this issue, the method proposed here replaced the binary discrimination between activations and noise by a more flexible shrinkage approach, so that morphology of the detected LAAs was intactly preserved by a hard thresholding and the undetected activations were denoised by a soft thresholding, instead of being directly removed. Additionally, for a better identification of LAAs an adaptive threshold, $\lambda_j(n)$, was computed for each scale, unlike the fixed cut-off used in~\cite{Martinez2017-cinc}. 
 Consequently, the novel SWT-based algorithm is able to remove most PLI in CFAEs at the reasonable cost of losing some low-frequency details in the LAAs. \par

\subsection{Limitations and further work}

PLI suppression has been mainly analyzed in a separate way from other common perturbations, such as baseline wandering and high-frequency noise. It is the best way to get an accurate assessment on the performance of the proposed SWT-based algorithm when dealing with PLI reduction. Nonetheless, the method's ability to reduce simultaneously the most relevant interferences in real EGMs will be addressed in future studies. To this respect, the PLI along with other nuisance signals have been successfully removed from the surface ECG recording~\cite{Xu2004,Bcharri2017,Lenis2017}.   
Hence, the development of an unified SWT-based methodology to obtain a compact, simple and computationally light preprocessing of EGM recordings also seems to be feasible for real-time implementations. Indeed, although wavelet-based algorithms require a higher computational time (i.e., 3$\mathrm{\mu}$s and 11.5$\mathrm{\mu}$s per sample for the SWT- and DWT-based methods, respectively) than those based on regular filtering (i.e., 70ns and 150ns per sample for fixed-bandwidth and adaptive filters, respectively), they can nowadays run easily online on specific digital signal processors or complex programmable logic devices~\cite{Gutierrez2017,Chen2015,Bahoura2012}.    
\par
Finally, although unipolar EGM recordings also play a key role in catheter ablation procedures~\cite{Issa2012}, the proposed SWT-based algorithm has not been validated over these signals. Despite the simplicity and versatility of the method, its application to these signals requires a thoughtful customization process. In fact, unipolar EGM recordings present different morphology and, due to the very dissimilar distance between recording electrodes, are disturbed by the PLI in a significantly different way than bipolar ones, thus involving some changes in the algorithm for a successful performance.

\section{Conclusions}\label{sec:conclusions}
A novel method based on the stationary wavelet transform has been proposed to reduce the powerline interference from intra-atrial and bipolar EGM recordings. The algorithm is characterized by its extreme simplicity as well as immunity to changes in amplitude and frequency of the interfering components. Moreover, unlike common fixed-bandwidth and adaptive notch filters, the proposed method avoids significant attenuation in the signal at the interference frequencies, thus reaching a widely better trade-off between noise reduction and original signal morphology preservation. As a consequence, the use of this algorithm in routine cardiac electrophysiology studies may be helpful for dealing with less disturbed EGM recordings, thus enabling possible revelation of new insights about the mechanisms triggering and maintaining AF. Nonetheless, some additional studies dealing with real EGM recordings with several simultaneous interferences will be required in the future.  
 
\section*{Acknowledgements}
Research supported by the grants DPI2017--83952--C3 MINECO/AEI/FEDER, UE and SBPLY/17/180501/000411 from Junta de Comunidades de Castilla-La Mancha
\section*{Conflicts of Interest}
The authors declare that they have no conflict of interest.

\section*{References}

\bibliographystyle{dcu}

\bibliography{20170509-mmartinez-PLIWT}   
\end{document}